\documentclass[aps, showpacs, twocolumn, pre, superscriptaddress, 10pt]{revtex4-1}
\usepackage{graphicx}
\usepackage{dcolumn}
\usepackage{color}
\usepackage{amssymb}
\usepackage{amsmath}
\usepackage{bm}
\usepackage{textcomp}
\usepackage{tabularx}
\usepackage{hyperref}
\usepackage{natbib} 
\usepackage[normalem]{ulem}
\usepackage{cancel}
\usepackage[usenames,dvipsnames]{xcolor}
\hypersetup{
colorlinks,
citecolor=blue,
filecolor=blue,
linkcolor=blue,
urlcolor=blue}
\begin{document}
\title{Dynamics of large-scale quantities in Rayleigh-B\'{e}nard convection}

\author{Ambrish Pandey}
\email{ambrishiitk@gmail.com}
\author{Abhishek Kumar}
\email{abhkr@iitk.ac.in}
\author{Anando G. Chatterjee}
\email{anandogc@iitk.ac.in}
\author{Mahendra K. Verma}
\email{mkv@iitk.ac.in}
\affiliation{Department of Physics, Indian Institute of Technology, Kanpur 208016, India}

\begin{abstract}
In this paper we estimate the relative strengths of various terms of the Rayleigh-B\'{e}nard equations.  Based on these estimates  and scaling analysis, we derive a general formula for the large-scale velocity, $U$, or the P\'{e}clet number that is applicable for arbitrary Rayleigh number $\mathrm{Ra}$ and Prandtl number $\mathrm{Pr}$.  Our formula fits reasonably well with the earlier simulation and experimental results.  Our analysis also shows that  the wall-bounded convection  has enhanced viscous force compared to free turbulence.  We also demonstrate how correlations  deviate the Nusselt number scaling from the theoretical prediction of $\mathrm{Ra}^{1/2}$ to the experimentally observed scaling of nearly $\mathrm{Ra}^{0.3}$.
\end{abstract}


\maketitle

\section{Introduction}

Modelling the  large-scale quantities in a turbulent flow is very important in many applications, e.g., fluid and magnetohydrodynamic turbulence, Rayleigh B\'{e}nard convection (RBC), rotating turbulence, etc~\cite{Pope:Book,Davidson:book2004,Lesieur:book,Frisch:Book,Davidson:book2013} . One such quantity is the magnitude of the large-scale velocity  that plays a critical role in physical processes. In this paper we will quantify this and other related quantities for RBC.

RBC is an idealized version of thermal convection in which a thin layer of fluid confined between two horizontal plates separated by a distance $d$ is heated from below and cooled from top.  The properties of RBC is specified using  two nondimensional parameters: the Rayleigh number $\mathrm{Ra}$, which is the ratio of the buoyancy term and the diffusion term, and the Prandtl number $\mathrm{Pr}$, which is the ratio of kinematic viscosity $\nu$ and thermal diffusivity $\kappa$~\cite{Ahlers:RMP2009, Siggia:ARFM1994, Lohse:ARFM2010, Chilla:EPJE2012}. Some of the important quantities of interest in RBC are the large-scale velocity $U$ or P\'{e}clet number $\mathrm{Pe} = U d/\kappa$, and the Nusselt number $\mathrm{Nu}$, which is the ratio of the total heat flux to the conductive heat flux. Note that the Reynolds number $\mathrm{Re} = \mathrm{Pe} /\mathrm{Pr}$. 

The P\'{e}clet and Nusselt numbers  are strong functions of  $\mathrm{Pr}$, as shown by Grossmann and Lohse~\cite{Grossmann:JFM2000, Grossmann:PRL2001, Grossmann:PRE2002, Stevens:JFM2013} (referred to as GL).  According to GL scaling, for small and moderate  $\mathrm{Pr}$'s, $\mathrm{Pe} \sim \sqrt{\mathrm{Ra}\mathrm{Pr}}$, but for large $\mathrm{Pr}$, $\mathrm{Pe \sim Ra^{3/5}}$.  The above scaling have been verified in many experiments~\cite{Xin:PRE1997, Cioni:JFM1997, Qiu:PRL2001, Ahlers:PRL2001, Niemela:JFM2001,Lam:PRE2002,Urban:PRL2012,He:PRL2012} and numerical simulations~\cite{Camussi:POF1998, Reeuwijk:PRE2008, Silano:JFM2010, BailonCuba:JFM2010, Scheel:JFM2012, Verma:PRE2012, Wagner:POF2013, Pandey:PRE2014, Horn:JFM2013}.  In this paper we derive a general formula for $\mathrm{Pe}$, of which the aforementioned relations are limiting cases, by comparing the relative strengths of the nonlinear, pressure, buoyancy, and viscous terms, and quantifying them using the numerical data.  Our derivation of the $\mathrm{Pe}$ formula  differs  from the Grossmann and Lohse's~\cite{Grossmann:JFM2000, Grossmann:PRL2001, Grossmann:PRE2002, Stevens:JFM2013} formalism; comparison between the two approaches will be discussed towards the end of the paper.

\begin{figure}
\includegraphics[scale=1]{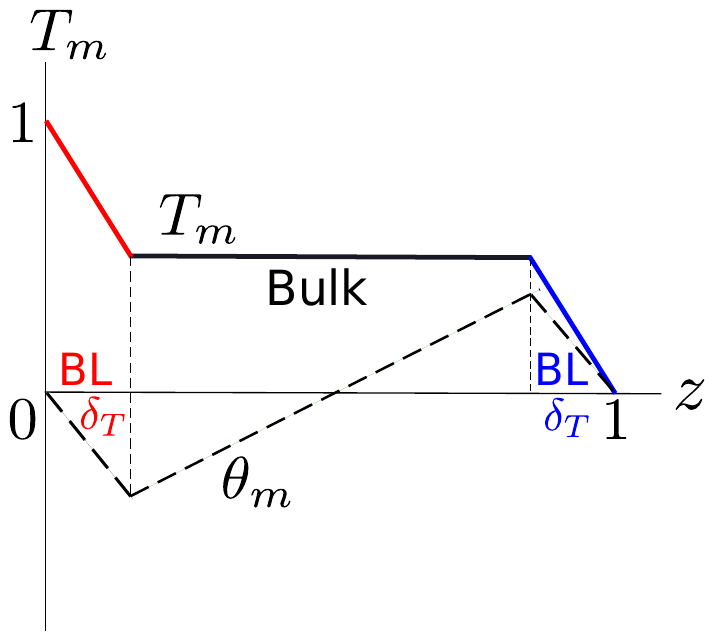}
\caption{ The plane-averaged temperature profile $T_m(z)$  exhibits  a near constant value in the bulk, and steep variations in the thermal boundary layers whose thickness is $\delta_T$.  The normalized temperature fluctuation $\theta = T+z-1$; the figure also exhibits its plane-averaged profile $\theta_m(z)$.}
\label{fig:Tz}
\end{figure}

Before describing the P\'{e}clet number formula, we briefly discuss the properties of the temperature fluctuations.  We normalize the temperature by the temperature difference between the plates $\Delta$, and the vertical coordinate $z$ by the plate distance $d$.  The profile of the plane-averaged temperature, $T_m(z)$, shown in Fig.~\ref{fig:Tz}, exhibits an approximate constant value of 1/2 in the bulk and steep variations in the thermal boundary layer.  It is customary to write RBC equations in terms of the temperature fluctuation from the conduction state,  $\theta(x,y,z)$, where $T = T_c(z) +\theta(x,y,z)$ with $T_c=1-z$ as the temperature profile for the conduction state.  The planar mean  of $\theta$, $\theta_m(z)$, is illustrated in Fig.~\ref{fig:Tz}.  

\begin{figure}
\includegraphics[scale=1]{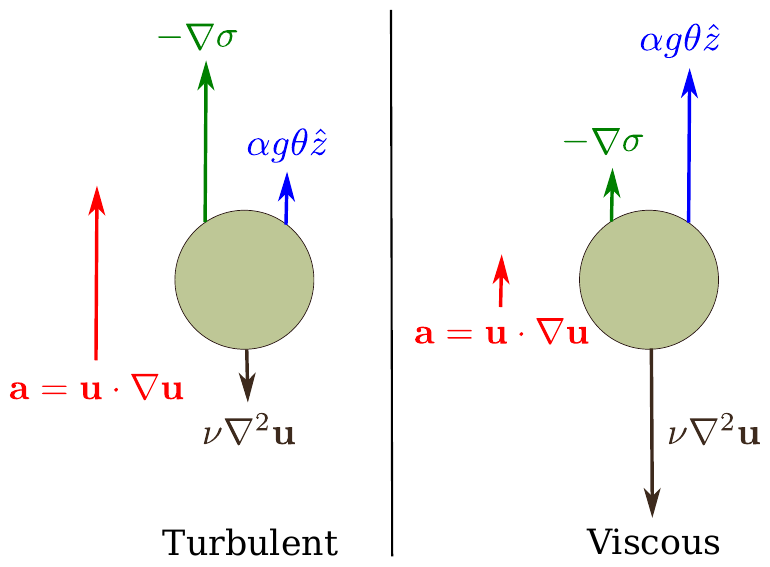}
\caption{Schematic  depiction of acceleration ${\bf a}$ of a fluid parcel that is rising against gravity: (a) in the turbulent regime, ${\bf a}$ is provided primarily by the pressure gradient $-\nabla \sigma$; (b) in the viscous regime, ${\bf a} \approx 0$ with the buoyancy and viscous terms balancing each other. The direction of the forces are reversed for  a descending fluid parcel. Note that buoyancy acts downward for the fluid parcels colder than the background.}  
\label{fig:schematic} 
\end{figure}

The momentum equation for  RBC is 
\begin{equation}
\partial_t {\bf u} + {\bf u} \cdot \nabla {\bf u}=  - \nabla \sigma + \alpha g \theta \hat{z} + \nu \nabla^2  {\bf u},  \label{eq:u}
\end{equation}
where ${\bf u}$ and $\sigma$ are the velocity and  pressure fields respectively,  $\alpha$ is the thermal expansion coefficient, and $g$ is the acceleration due to gravity. Here we have taken the density to be unity.   Under a steady state ($\langle \partial_t {\bf u} \rangle = 0$), the acceleration of a fluid parcel, ${\bf u} \cdot \nabla {\bf u}$, is provided by the pressure gradient $- \nabla \sigma$, buoyancy $\alpha g \theta \hat{z}$, and the viscous term  $\nu \nabla^2  {\bf u}$.  For a viscous flow, the net force or acceleration is very small, and the viscous force balances the buoyancy. However, in the turbulent regime, the pressure gradient dominates the buoyancy and viscous forces  (see Fig.~\ref{fig:schematic}).  We quantify these accelerations using numerical data that help us understand the scaling of the P\'{e}clet number and related quantities for arbitrary $\mathrm{Pr}$ and $\mathrm{Ra}$. The velocity Fourier modes $\hat{\bf u}(0,0,k_z) = 0$ due to incompressibility and no-mean flow condition.  Hence $\hat{\theta}_m(0,0,k_z)$ corresponding to $\theta_m(z)$ follows a relation: $i k_z \hat{\sigma}_m(0,0,k_z) = \alpha g \hat{\theta}_m(0,0,k_z) $ or $d \sigma_m(z) /dz = \alpha g \theta_m$.  Therefore, in the momentum equation, the Fourier modes other than $(0,0,k_z)$ involve  $\theta_\mathrm{res} = \theta-\theta_m$ and $\sigma_\mathrm{res} = \sigma - \sigma_m$ respectively (see Appendix). 

\section{Numerical Method}

We solve the RBC equations~\cite{Ahlers:RMP2009, Siggia:ARFM1994, Lohse:ARFM2010, Chilla:EPJE2012} in a three-dimensional unit box for $\mathrm{Pr} = 1, 6.8, 10^2, 10^3$ and Rayleigh numbers between $10^6$ and   $5 \times 10^8$ on grids $60^3, 80^3, 100^3$ and $256^3$ using the finite volume code OpenFOAM~\cite{OpenFOAM}. We employ the no-slip boundary condition for the velocity field at all the walls, the conducting boundary condition at the top and bottom walls, and the insulating boundary condition at the vertical walls. For time stepping we use second-order Crank-Nicolson scheme.  To resolve the boundary layers we employ a nonuniform mesh with higher concentration of grid points (greater than $4$\textendash$6$) near the boundaries~\cite{Grotzbach:JCP1983, Shishkina:NJP2010}. Table~\ref{table:details} includes the summary of numerical simulations performed. We have performed the grid-independence test for $\mathrm{Pr = 1, Ra = 10^8}$ by performing simulations on $100^3$ and $256^3$ grids, and find that the Nusselt and P\'{e}clet numbers are different by approximately 3\% and 1\%, respectively.
\begin{table*}
\begin{ruledtabular}
\caption{Details of our numerical simulations performed in a unit box: the Prandtl number ($\mathrm{Pr}$), the Rayleigh number ($\mathrm{Ra}$),  grid points ($N^3$), the Nusselt number (Nu), the P\'{e}clet number (Pe),  averaged values of the nonlinear term, pressure gradient, buoyancy, and viscous force, and the number of snapshots over which the time-averaging has been performed.  The terms of the momentum equation are normalized by $Z = \alpha g \Delta$.}
\begin{tabular}{cccccccccc}
$\mathrm{Pr}$ & $\mathrm{Ra}$ & $N^3$ & Nu & Pe & $|{\bf u} \cdot \nabla {\bf u}|/Z$ & $|(-\nabla \sigma)_{\mathrm{res}}|/Z$ & $|\alpha g \theta \hat{z}|/Z$ & $|\nu \nabla^2 {\bf u}|/Z$ & snapshots \\ \hline
1 & $1 \times 10^6$ & $60^3$ & 8.0 & 146.1  & 0.152 & 0.110 & 0.0955 & 0.110  & 100  \\
1 & $2 \times 10^6$ & $60^3$ & 10.0 & 211.3 & 0.177 & 0.126 & 0.0883 & 0.0934 & 200  \\
1 & $5 \times 10^6$ & $60^3$ & 13.4 & 340.3 & 0.213 & 0.146 & 0.0788 & 0.0749 & 200  \\
1 & $1 \times 10^7$ & $80^3$ & 16.3 & 485.4 & 0.234 & 0.157 & 0.0715 & 0.0707 & 50   \\
1 & $2 \times 10^7$ & $80^3$ & 20.2 & 687.4 & 0.266 & 0.168 & 0.0653 & 0.0608 & 200  \\
1 & $5 \times 10^7$ & $80^3$ & 26.8 & 1103 &  0.318 & 0.187 & 0.0575 & 0.0486 & 178  \\
1 & $1 \times 10^8$ & $100^3$ & 32.9 & 1554 & 0.359 & 0.200 & 0.0526 & 0.0480 & 100  \\
1 & $1 \times 10^8$ & $256^3$ & 31.9 & 1537 & 0.348 & 0.205 & 0.0529 & 0.0789 & 35 \\
1 & $5 \times 10^8$ & $256^3$ & 51.2 & 3408 & 0.472 & 0.233 & 0.0429 & 0.0559 & 34 \\
6.8 & $1 \times 10^6$ & $60^3$ & 8.4 & 182.7  & 0.0273 & 0.0160 & 0.0825 & 0.106  & 200  \\
6.8 & $2 \times 10^6$ & $60^3$ & 9.9 & 252.8  & 0.0333 & 0.0221 & 0.0745 & 0.0865 & 200  \\
6.8 & $5 \times 10^6$ & $60^3$ & 13.1 & 413.6 & 0.0413 & 0.0272 & 0.0645 & 0.0691 & 250  \\
6.8 & $1 \times 10^7$ & $80^3$ & 16.2 & 608.6 & 0.0474 & 0.0310 & 0.0581 & 0.0669 & 163  \\
6.8 & $2 \times 10^7$ & $80^3$ & 20.3 & 903.2 & 0.0558 & 0.0352 & 0.0518 & 0.0570 & 200  \\
6.8 & $5 \times 10^7$ & $80^3$ & 27.7 & 1536  & 0.0696 & 0.0419 & 0.0452 & 0.0456 & 165  \\
$10^2$ & $1 \times 10^6$ & $60^3$ & 8.5 & 190.7  & $1.94 \times 10^{-3}$ & $1.02 \times 10^{-3}$ & 0.0792 & 0.108 & 300 \\
$10^2$ & $2 \times 10^6$ & $60^3$ & 11.2 & 278.2 & $2.19 \times 10^{-3}$ & $1.01 \times 10^{-3}$ & 0.0731 & 0.0938 & 180 \\
$10^2$ & $5 \times 10^6$ & $60^3$ & 14.5 & 500.0 & $2.91 \times 10^{-3}$ & $1.42 \times 10^{-3}$ & 0.0587 & 0.0732 & 250 \\
$10^2$ & $1 \times 10^7$ & $80^3$ & 17.1 & 704.2 & $3.69 \times 10^{-3}$ & $2.15 \times 10^{-3}$ & 0.0531 & 0.0659 & 250 \\
$10^2$ & $2 \times 10^7$ & $80^3$ & 20.7 & 1044  & $4.47 \times 10^{-3}$ & $2.62 \times 10^{-3}$ & 0.0467 & 0.0556 & 200 \\
$10^2$ & $5 \times 10^7$ & $80^3$ & 27.7 & 1826  & $6.08 \times 10^{-3}$ & $3.53 \times 10^{-3}$ & 0.0395 & 0.0460 & 250 
\end{tabular}
\label{table:details}
\end{ruledtabular} 
\end{table*}

\begin{figure}
\includegraphics[scale=0.95]{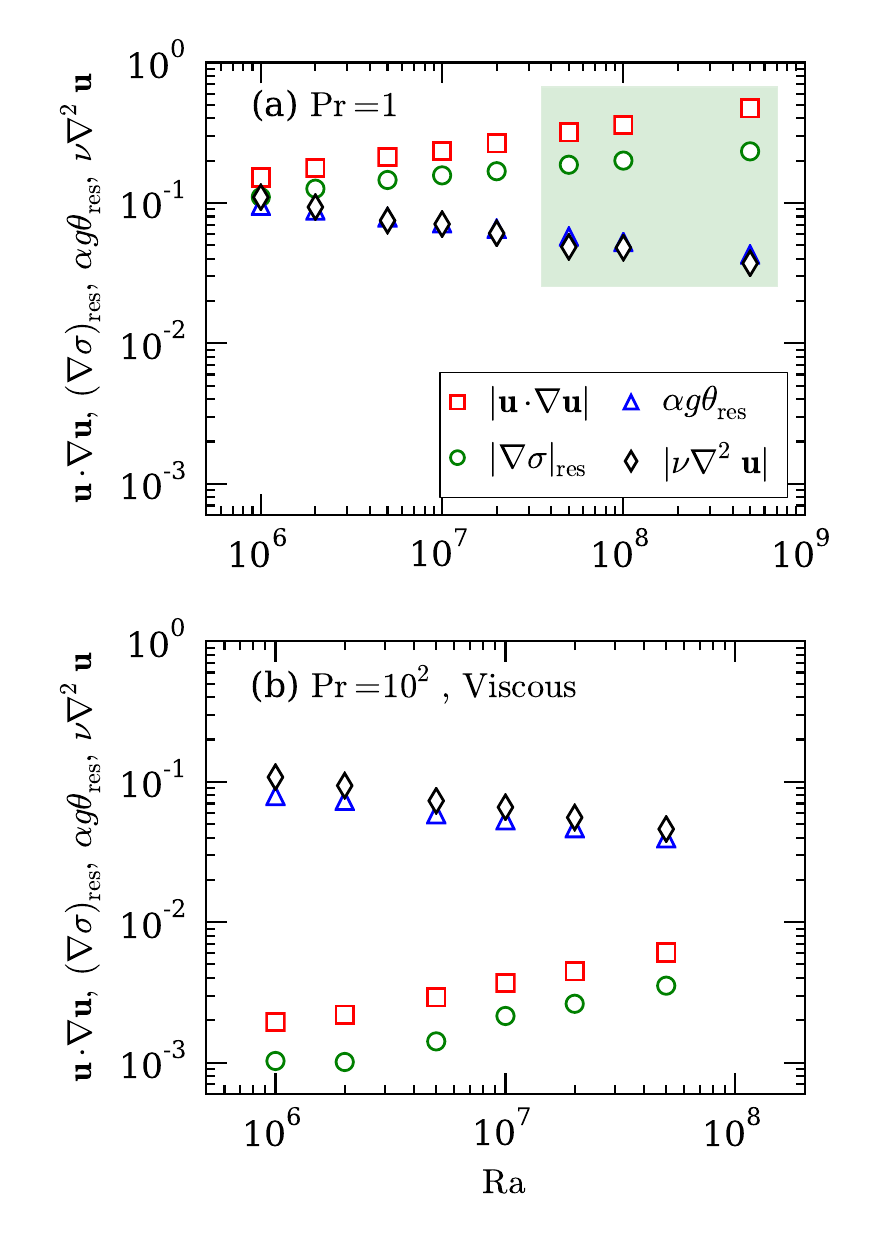}
\caption{Plots comparing the root mean square (rms) values of various terms of the momentum equation, $|{\bf u} \cdot \nabla {\bf u|}$, $|(\nabla \sigma)_{\mathrm{res}}|$, $\alpha g \theta_{\mathrm{res}}$, and $|\nu \nabla^2 {\bf u}|$ as function of $\mathrm{Ra}$ for (a) $\mathrm{Pr} = 1$, and (b) $\mathrm{Pr} =  10^2$.  In the turbulent regime (the shaded region of Fig.~(a)), $|{\bf u} \cdot \nabla {\bf u}| \approx |\nabla \sigma| \gg \alpha g \theta_\mathrm{res}, |\nu \nabla^2 {\bf u}|$; in the viscous regime (Fig.~(b)), $\alpha g \theta_\mathrm{res}, |\nu \nabla^2 {\bf u}| \gg |{\bf u} \cdot \nabla {\bf u}|, |\nabla \sigma|$.}
\label{fig:terms}
\end{figure}
We compute $U = \sqrt{\langle u^2 \rangle}$ (rms value)   first by volume averaging $u^2 $ over the entire computational domain, and then by performing temporal average over many snapshots (see Table~\ref{table:details}).  We  also estimate the strengths of various terms of the momentum equation by  similar averaging process.  These values are depicted in Fig.~\ref{fig:terms}(a,b) for $\mathrm{Pr}=1$ and $10^2$ with $ \mathrm{Ra}$ ranging from $10^6$ to $5\times 10^8$.  The flow is turbulent for $\mathrm{Pr}=1$ and  $ \mathrm{Ra} = 5 \times 10^7, 10^8, 5\times 10^8$ whose respective Reynolds numbers are approximately 1103, 1537, and 3408.  But the flow is laminar for all $ \mathrm{Ra}$'s when $\mathrm{Pr}=10^2$.  Clearly, the numerical values are consistent with the schematic diagram of Fig.~\ref{fig:schematic}.  The acceleration in the turbulent regime is dominated by the pressure gradient, with small contributions from the buoyancy and viscous terms.  However, in the viscous regime, the viscous term balances the buoyancy yielding a very small acceleration. Therefore, all the terms of the momentum equation balance reasonably well. To test this out, we compute the vector sum of all the terms for a given Ra, i.e., $S = |{\bf u} \cdot \nabla {\bf u}| - |(-\nabla \sigma)_{\mathrm{res}}| - |\alpha g \theta_{\mathrm{res}} \hat{z}| + |\nu \nabla^2 {\bf u}|$, and the most dominant term among them, i.e., $M = \mathrm{Maximum}(|{\bf u} \cdot \nabla {\bf u}|, |(-\nabla \sigma)_{\mathrm{res}}|, |\alpha g \theta_{\mathrm{res}} \hat{z}|, |\nu \nabla^2 {\bf u}|)$. We observe that the ratio $S/M$ varies from 19\% to 53\%. We also remark that the buoyancy  in RBC is $\alpha g (T - T_{\mathrm{ref}}) \hat{z}$. The temperature $T$ of a cold  fluid parcel (moving downward) is lower than that of the reference fluid temperature $T_{\mathrm{ref}}$. Therefore the force due to the density difference on a colder fluid parcel traveling downwards is in $- \hat{z}$ direction, and hence the forces on it are opposite to that in Fig.~\ref{fig:schematic}. Thus, the net force balance holds for both ascending and descending fluid parcels. 

\section{Results}

\begin{figure}
\includegraphics[scale=1]{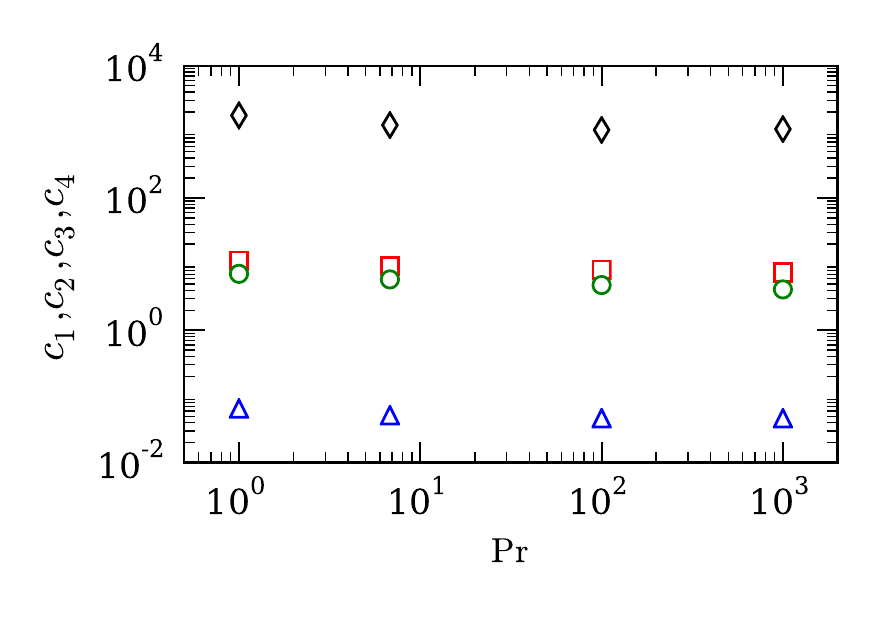}
\caption{Plot of $c_1$ (red squares), $c_2$ (green circles), $c_3$ (blue triangles), and $c_4$ (black diamonds) of Eqs.~(\ref{eq:c_1}--\ref{eq:c_4}) as function of $\mathrm{Pr}$ for $\mathrm{Ra} = 2 \times 10^7$. All $c_i$'s are weak functions of  $\mathrm{Pr}$.}
\label{fig:c_pr}
\end{figure}

To quantify the terms of the momentum equation, we perform a scaling analysis of the momentum equation that yields
\begin{equation}
c_1 \frac{U^2}{d} =  c_2  \frac{U^2}{d}+  c_3 \alpha g \Delta  - c_4 \nu \frac{U}{d^2}, \label{eq:U}
\end{equation}
where $ c_1 =  |{\bf u \cdot \nabla u}| / (U^2/d)$, $c_2 = |(\nabla \sigma)_{\mathrm{res}}| / (U^2/d)$, $c_3  =  \theta_{\mathrm{res}}/\Delta$, and $c_4 = |\nabla^2 {\bf u}| /(U/d^2)$ are dimensionless coefficients.    In this paper, we compute the  coefficients $c_i$'s for $\mathrm{Pr} = 1, 6.8, 100$, and 1000. Our numerical computation yields
\begin{subequations}
\begin{eqnarray}
c_1 & = & 1.5 \mathrm{Ra}^{0.10} \mathrm{Pr}^{-0.06},   \label{eq:c_1}\\
c_2 & = & 1.6 \mathrm{Ra}^{0.09} \mathrm{Pr}^{-0.08},  \label{eq:c_2}\\
c_3 & = & 0.75 \mathrm{Ra}^{-0.15} \mathrm{Pr}^{-0.05},  \label{eq:c_3}\\
c_4 & = & 20 \mathrm{Ra}^{0.24} \mathrm{Pr}^{-0.08}. \label{eq:c_4}
\end{eqnarray}	
\end{subequations}
The errors in the exponents are $\lessapprox 0.01$, whereas the prefactors are uncertain by approximately 10\%. However in the $c_4$ vs.~$\mathrm{Ra}$ scaling, the exponent 0.24 is uncertain by approximately 30\%. Clearly the coefficients are weak functions of $\mathrm{Pr}$ (see Fig.~\ref{fig:c_pr}), but they show significant variations with $\mathrm{Ra}$. This aspect is in contrast to  {\em unbounded flows}  (without walls) where the coefficients are independent of parameters. We attribute the above scaling to the thermal plates or {\em bounded flow}.  Note that the $\mathrm{Ra}$ dependence of $c_4$  leads to an enhanced viscous force for RBC  compared to  free turbulence. In our simulations, $1 \le \mathrm{Pr} \le 10^2$ and $\mathrm{Ra} \le 5\times 10^8$, hence $c_i$'s of Eqs.~(\ref{eq:c_1}-\ref{eq:c_4}) may get altered for larger $\mathrm{Ra}$ and extreme $\mathrm{Pr}$.  Also, $c_i$'s should depend, at least weakly, on geometry and aspect ratio.  Yet we believe that our formula, to be described below, should provide approximate description of $U$ for large $\mathrm{Ra}$ and extreme $\mathrm{Pr}$'s.

\begin{figure}
\includegraphics[scale=1]{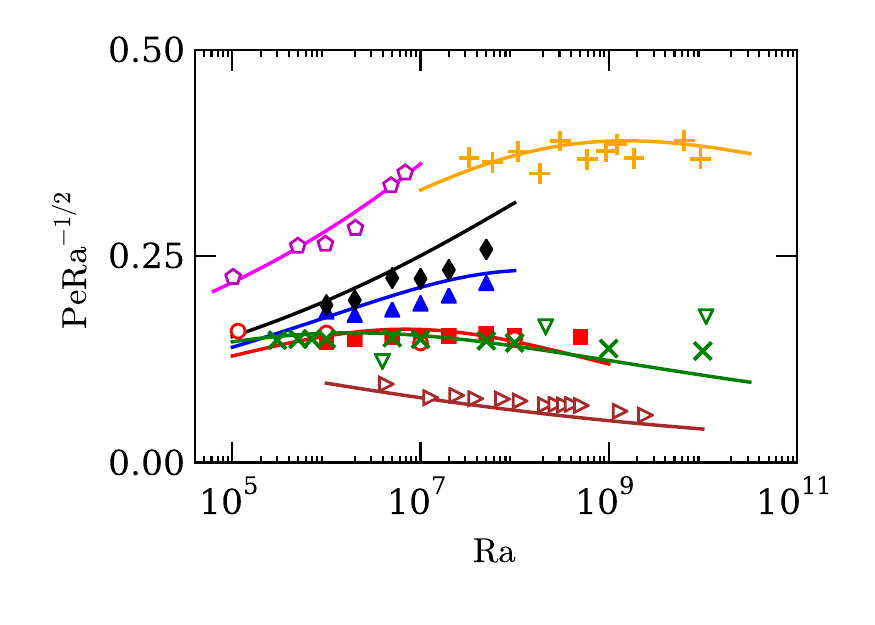}
\caption{Plot of normalized P\'{e}clet number $\mathrm{PeRa^{-1/2}}$ as a function of $\mathrm{Ra}$ for different $\mathrm{Pr}$. Our formula for P\'{e}clet number $\mathrm{Pe}$ [Eq.~(\ref{eq:Pe_analy})] (shown as continuous curves) fits reasonably well with the numerical results of ours  ($\mathrm{Pr} = 1$, red squares; $\mathrm{Pr} = 6.8$, blue triangles;  $\mathrm{Pr} = 10^2$, black diamonds), Scheel and Schumacher~\cite{Scheel:JFM2014} ($\mathrm{Pr} = 0.7$, green crosses), Reeuwijk \textit{et al.}~\cite{Reeuwijk:PRE2008}  ($\mathrm{Pr} = 1$, red  circles), and Silano \textit{et al.}~\cite{Silano:JFM2010} ($\mathrm{Pr} = 10^3$, magenta	 pentagons), and the experimental results of Cioni \textit{et al.}~\cite{Cioni:JFM1997}  ($\mathrm{Pr} \approx 0.022$, brown right-triangles) and Xin and Xia~\cite{Xin:PRE1997}  ($\mathrm{Pr} \approx 6.8$, orange pluses).  The agreement is deficient for Niemela \textit{et al.}'s~\cite{Niemela:JFM2001} experimental result ($\mathrm{Pr} \approx 0.7$, green down-triangles), possibly due to various factors discussed in the text. }
\label{fig:pe}
\end{figure}

Multiplication of Eq.~(\ref{eq:U}) with $d^3/\kappa^2$ yields the following equation for the  P\'{e}clet number:
\begin{equation}
c_1 \mathrm{Pe}^2 = c_2 \mathrm{Pe}^2 + c_3 \mathrm{RaPr} - c_4 \mathrm{PePr}, \label{eq:Pe}
\end{equation}
whose  solution is
\begin{equation}
\mathrm{Pe} = \frac{-c_4 \mathrm{Pr} + \sqrt{c_4^2 \mathrm{Pr}^2 + 4 c_3(c_1-c_2)\mathrm{RaPr}}}{2 (c_1-c_2)}. \label{eq:Pe_analy}
\end{equation}	
A combination of Eqs.~(\ref{eq:Pe_analy}, \ref{eq:c_1}-\ref{eq:c_4}) provide us a predictive formula for the P\'{e}clet number for arbitrary $\mathrm{Pr}$ and $\mathrm{Ra}$.  We test our formula with  numerical results of ours, Reeuwijk \textit{et al.}~\cite{Reeuwijk:PRE2008}, Silano \textit{et al.}~\cite{Silano:JFM2010}, and Scheel and Schumacher~\cite{Scheel:JFM2014}, and the experimental results of Niemela {\em et al.}~\cite{Niemela:JFM2001}, Xin and Xia~\cite{Xin:PRE1997} and Cioni \textit{et al.}~\cite{Cioni:JFM1997}.  The predictions of Eq.~(\ref{eq:Pe_analy}) for $\mathrm{Pr} = 0.022$ and Pr = 6.8 have been multiplied with 2.5 and 1.2, respectively, to fit the experimental results from Cioni {\em et al.}~\cite{Cioni:JFM1997}  and Xin and Xia~\cite{Xin:PRE1997}.   As shown in Fig.~\ref{fig:pe}, our formula describes the numerical and experimental data reasonably well for Prandtl numbers ranging from 0.025 to 1000 and for various geometries.  However the above factors  (2.5 and 1.2) and the discrepancy between our predictions and the results of Niemela {\em et al.}~\cite{Niemela:JFM2001} are due to the aforementioned uncertainty in $c_i$'s, geometrical factors, aspect ratio dependence, and different definitions used for $U$. 

The limiting cases of $\mathrm{Pe}$ formula of  Eq.~(\ref{eq:Pe_analy}) along with Eqs.~(\ref{eq:c_1}-\ref{eq:c_4}) yield {\em turbulent} and {\em viscous} regimes.   For $c_4^2 \mathrm{Pr}^2  \ll 4 c_3(c_1-c_2)\mathrm{RaPr} $ or $\mathrm{Ra} \gg 10^6  \mathrm{Pr}$, we obtain the turbulent regime with
 \begin{equation}
\mathrm{Pe} = \sqrt{\frac{c_3}{|c_1-c_2|} \mathrm{RaPr}} \approx \sqrt{7.5  \mathrm{Pr}}\mathrm{Ra}^{0.38}.
\end{equation} 
For mercury ($\mathrm{Pr} = 0.025$), the above formula yields $(\mathrm{Pe})_{\mathrm{theory}}= 0.38\mathrm{Ra}^{0.38}$, which is in a reasonable agreement with Cioni {\em et al.}'s~\cite{Cioni:JFM1997} finding that  $(\mathrm{Pe})_{\mathrm{expt}}= 0.24 \mathrm{Ra}^{0.43}$ in their  convection experiment with mercury.  However, for $\mathrm{Pr}\approx 1$ and large $\mathrm{Ra}$, many authors~\cite{Xin:PRE1997, Niemela:JFM2001, Qiu:PRL2001,Camussi:POF1998, Reeuwijk:PRE2008, Scheel:JFM2012, Scheel:JFM2014, Verma:PRE2012, Stevens:JFM2013} as well our numerical simulation show that $\mathrm{Pe} \sim \mathrm{Ra}^{0.51}$, not $\mathrm{Pe} \sim \mathrm{Ra}^{0.38}$.  This is because the condition  $\mathrm{Ra} \gg 10^6 \mathrm{Pr}$ is not satisfied in these systems (except for~\cite{Niemela:JFM2001} and~\cite{Scheel:JFM2014}). We remark that Eqs.~(\ref{eq:c_1}-\ref{eq:c_4}) and the condition $\mathrm{Ra} \gg 10^6 \mathrm{Pr}$ are derived from our numerical simulations for moderate values of $\mathrm{Ra}$ and $\mathrm{Pr}$; these equations may change somewhat  for larger Rayleigh and Prandtl numbers.  Aspect ratio and geometry could also affect the scaling of $c_i$'s. Yet, our $\mathrm{Pe}$ formula describes the earlier experimental and numerical simulations reasonably well (see Fig.~\ref{fig:pe}).

The other limiting case (viscous regime), $c_4^2 \mathrm{Pr}^2  \gg 4 c_3(c_1-c_2)\mathrm{RaPr} $, yields $\mathrm{Pe} = (c_3/c_4) \mathrm{Ra} \approx 0.0375 \mathrm{Ra}^{0.61}$, which is independent of $\mathrm{Pr}$ as observed in several numerical simulations and  experiments~\cite{Lam:PRE2002, Silano:JFM2010, Horn:JFM2013, Stevens:JFM2013, Pandey:PRE2014}. An application of the above to the Earth's mantle~\cite{Schubert:book2001, Turcotte:book2002, Galsa:SE2015} with parameters $d \approx 2900$ km, $\kappa \approx 10^{-6} \mathrm{m}^2/\mathrm{s}$, $\mathrm{Ra} \approx 5 \times 10^7$, $\mathrm{Pr} \approx 10^{23}-10^{24}$, and $U \approx 20$ mm/yr  yields $(\mathrm{Pe})_{\mathrm{obs}} \approx 1840$, which is  close to our prediction that $(\mathrm{Pe})_{\mathrm{theory}} \approx 1580$.

The effects of the thermal plates and the boundary layers come into play in several different ways in our  analysis.  Firstly, the thermal boundary layers at the two walls induce a mean temperature profile $\theta_m(z)$ (see Fig.~\ref{fig:Tz}), and only $\theta_\mathrm{res} = \theta - \theta_m(z)$ participate in the momentum equation.  Secondly, the boundary layers induce $\mathrm{Ra}$ dependence  on $c_i$'s  [see Eq.~(\ref{eq:c_1}-\ref{eq:c_4})].  Note that in unbounded flows  (without walls) the corresponding $c_i$'s  are expected to be constants, i.e.,  independent of system parameters like $\mathrm{Ra}$.  Also,  the ratio of the nonlinear term and the viscous term of Eq.~(\ref{eq:U}) is $(Ud/\nu) (c_1/c_4) \sim \mathrm{Re} \mathrm{Ra}^{-0.14}$, not just Reynolds number $\mathrm{Re}$ as in unbounded flows.  Thus, the thermal plates or the boundary layers enhance the dissipation in RBC flows compared to unbounded flows.  We show below that a similar behaviour is observed for the Nusselt number and the viscous dissipation rates.

The anisotropy induced by the thermal plates have important consequences on the heat transport and dissipation rates:  $\langle u_z \theta \rangle \ne 0$, unlike $\langle u_z \theta \rangle = 0$ for any isotropic flow. The Nusselt number $\mathrm{Nu} \sim  C_{u\theta_\mathrm{res}} \langle u_z^2 \rangle^{1/2} \langle \theta_\mathrm{res}^2 \rangle^{1/2}$, where $C_{u\theta_\mathrm{res}} = \langle u_z \theta_\mathrm{res} \rangle/[\langle u_z^2 \rangle^{1/2} \langle \theta_\mathrm{res}^2 \rangle^{1/2}] $ is the normalized correlation function. In Table~\ref{tab:Ra_dependence} we list the $\mathrm{Ra}$ dependence for  the above quantities that  provide the appropriate corrections to the Nusselt number from the Kraichnan's predictions~\cite{Kraichnan:POF1962}   for the ultimate regime ($ \mathrm{Nu}  \sim \mathrm{Ra}^{1/2}$) to the experimentally observed $ \mathrm{Nu}  \sim \mathrm{Ra}^{0.30}$.

It is tempting to connect our findings to the ultimate regime of turbulent convection~\cite{Urban:PRL2012,He:PRL2012}.  We conjecture that in the ultimate regime, $C_{u\theta_\mathrm{res}}$ and $\theta_\mathrm{res}$, as well as the coefficients $c_i$'s, would become independent of $\mathrm{Ra}$ due to boundary layer detachment, and hence yield  $ \mathrm{Nu}  \sim \mathrm{Ra}^{1/2}$.  We need inputs from experiments and numerical simulations to probe the above conjectures.

\begin{table}
\caption{ $\mathrm{Ra}$ dependence  of the normalized correlation function $C_{u\theta_\mathrm{res}} = \langle u_z \theta_\mathrm{res} \rangle/[\langle u_z^2 \rangle^{1/2} \langle \theta_\mathrm{res}^2 \rangle^{1/2}] $,   $\langle \theta_\mathrm{res}^2 \rangle^{1/2} $,   $\langle u_z^2 \rangle^{1/2} $, $\mathrm{Nu}$, and $\epsilon_u$ computed using numerical data. The data in the shaded region of Fig.~\ref{fig:terms}(a) for $\mathrm{Pr} = 1$ belongs to the turbulent regime whereas that of $\mathrm{Pr} = 10^2$ belongs to the viscous regime.  The functional forms of the table are based on our numerical simulations. The errors in the exponents are around 5\%--10\% for all the quantities.}
\begin{ruledtabular}
\begin{tabular}{ccc}

 & Turbulent regime & Viscous regime \\
\hline 
 
$C_{u \theta_\mathrm{res}}$ &  $\mathrm{Ra}^{-0.05}$   & $\mathrm{Ra}^{-0.07}$ \\

 $\langle \theta_\mathrm{res}^2 \rangle^{1/2} $  & $\mathrm{Ra}^{-0.13}$ &  $\mathrm{Ra}^{-0.18}$\\
 
  $\langle u_z^2 \rangle^{1/2} $  & $\mathrm{Ra}^{0.51}$ &  $\mathrm{Ra}^{0.58}$ \\
  
    $\mathrm{Nu}$  & $\mathrm{Ra}^{0.32}$ &  $\mathrm{Ra}^{0.33}$ \\
  
  $\epsilon_u$ & $(U^3/d) \mathrm{Ra}^{-0.21}$ &  $(\nu U^2/d^2)   \mathrm{Ra}^{0.17}$ \\

\end{tabular}
\end{ruledtabular}
\label{tab:Ra_dependence}
\end{table}

The exact relations of Shraiman and Siggia~\cite{Shraiman:PRA1990} yields
\begin{equation}
\epsilon_u  = \frac{U^3}{d}  \frac{\mathrm{(Nu-1)RaPr}}{\mathrm{Pe}^3}. \label{eq:eps_u1}
\end{equation}
In the turbulent regime, $ \mathrm{Nu} \sim \mathrm{Ra}^{0.32}$ and $ \mathrm{Pe} \sim \mathrm{Ra}^{0.51}$, hence, $\epsilon_u  \ne U^3/d$, rather $ \epsilon_u  \sim (U^3/d) \mathrm{Ra}^{-0.21}$ due the confinement of turbulence by the walls. In the viscous regime,
\begin{equation}
\epsilon_u  = \frac{\nu U^2}{d^2}  \frac{\mathrm{(Nu-1)Ra}}{\mathrm{Pe}^2}. \label{eq:eps_u2}
\end{equation}
Since  $ \mathrm{Nu} \sim \mathrm{Ra}^{0.33}$ and $ \mathrm{Pe} \sim \mathrm{Ra}^{0.58}$~\cite{Silano:JFM2010, Pandey:PRE2014}, we observe that  $\epsilon_u \sim (\nu U^2/d^2 ) \mathrm{Ra}^{0.17}$. Note that in a typical viscous scenario,   $\epsilon_u  \sim \nu U^2/d^2$.  Hence the $\mathrm{Ra}$ dependence of $c_4$ and the correction of the viscous term from $\nu U/d^2$ are related to the boundary layers. 
\begin{figure*}
\includegraphics[scale=0.7]{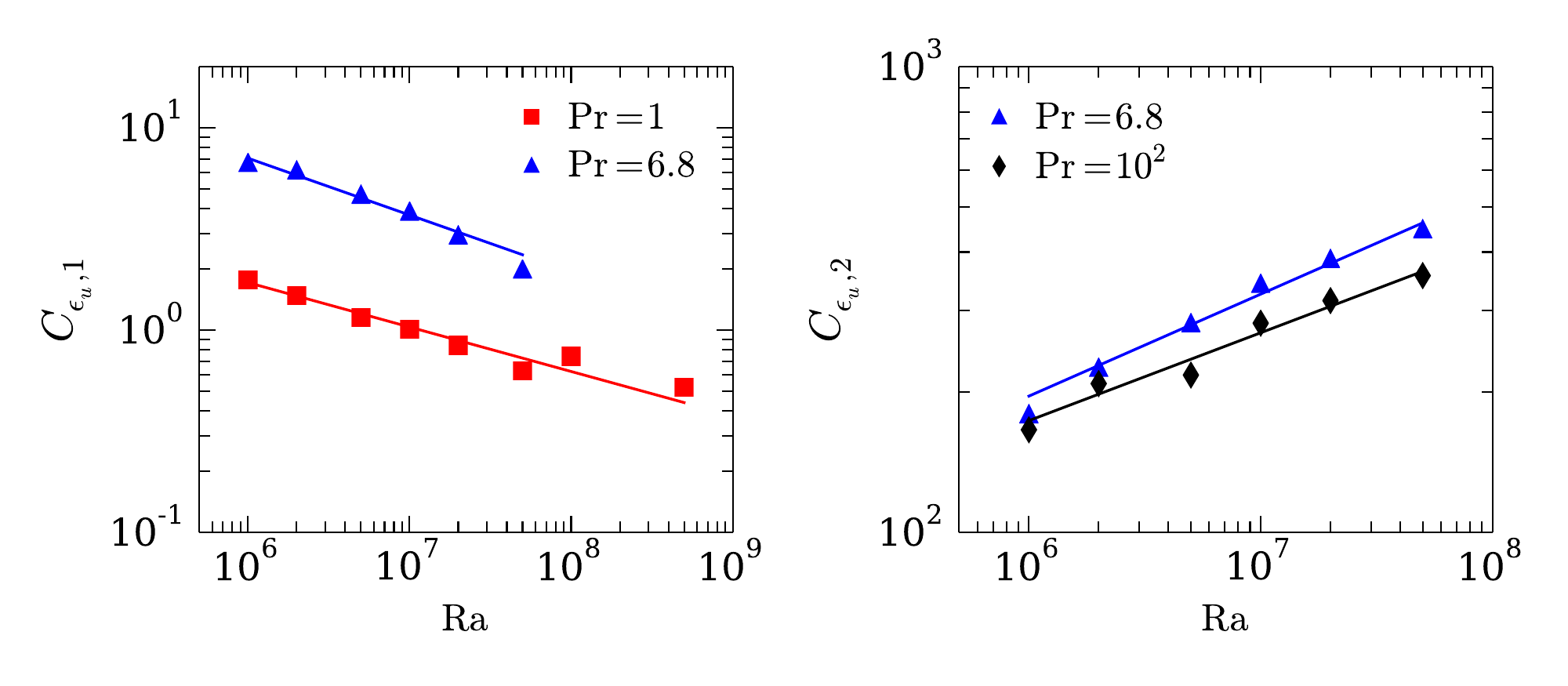}
\caption{The plots of normalized viscous dissipation rates $C_{\epsilon_u,1}$ (in turbulent regime) and  $C_{\epsilon_u,2}$ (in viscous regime). $C_{\epsilon_u,1} \sim \mathrm{Ra}^{-0.22}$ and $\mathrm{Ra}^{-0.25}$ for $\mathrm{Pr} = 1$ and $6.8$ respectively, while $C_{\epsilon_u,2} \sim \mathrm{Ra}^{0.22}$ and $\mathrm{Ra}^{0.19}$ for $\mathrm{Pr} = 6.8$ and $10^2$ respectively. These results are in qualitative agreement with model predictions (see Table~\ref{tab:Ra_dependence}). }
\label{fig:c_eps}
\end{figure*}

To quantify the scaling of the viscous dissipation rates given by Eqs.~(\ref{eq:eps_u1}, \ref{eq:eps_u2}), we define two normalized viscous dissipation rates as
\begin{eqnarray}
C_{\epsilon_u,1} & = & \frac{\epsilon_u}{U^3/d} = \frac{\mathrm{(Nu-1)RaPr}}{\mathrm{Pe}^3} \sim \mathrm{Ra^{-0.21} Pr},  \\
C_{\epsilon_u,2} & = & \frac{\epsilon_u}{\nu U^2/d^2} = \frac{\mathrm{(Nu-1)Ra}}{\mathrm{Pe}^2} \sim \mathrm{Ra}^{0.17}.
\end{eqnarray}
The correlation functions $C_{\epsilon_u,1}, C_{\epsilon_u,2}$ are suitable for the turbulent and viscous regimes respectively. We compute these quantities using the numerical data obtained from numerical simulations and plot them as a function of $\mathrm{Ra}$ in Fig.~\ref{fig:c_eps}, and demonstrate that $C_{\epsilon_u,1} \sim \mathrm{Ra^{-0.21}}$ and $C_{\epsilon_u,2} \sim  \mathrm{Ra}^{0.17}$.  Our results demonstrate that the dissipation rates in RBC differ from those in unbounded flows due to the walls.

Let us estimate the ratio of the total dissipation rates (product of the dissipation rate and the appropriate volume) in the boundary layer ($D_{u,BL}$) and in the bulk ($D_{u,bulk}$).  In the turbulent regime
\begin{eqnarray}
\frac{D_{u,BL}}{D_{u,bulk}} & \approx & \frac{(\epsilon_{u,BL}) (2 A \delta_u)}{(\epsilon_{u,bulk}) (A d - 2 A \delta_u)}  \approx  \left( \frac{2 \nu U^2/\delta_u^2}{U^3/d} \right) \frac{\delta_u}{d} \nonumber \\
&\approx &  2\frac{d/\delta_u}{\mathrm{Re}} \approx2 \mathrm{Re}^{-1/2}
\label{eq:Du}
\end{eqnarray}
since $\delta_u/d \sim \mathrm{Re}^{-1/2}$~\cite{Landau:FM1987}.  Here $\delta_u$ is the thickness of the viscous boundary layers at the top and bottom plates, and $A$ is the cross section area of these plates.  The factor 2 is included to account for the dissipation near both the plates.   Equation~(\ref{eq:Du}) indicates that $D_{u,BL} \ll D_{u,bulk}$ for large $\mathrm{Re}$.  In the viscous regime, the boundary layer spans the whole region ($2 \delta_u \approx d$), therefore $D_{u,BL}$ dominates $ D_{u,bulk}$. Also, our formula [Eq.~(\ref{eq:Pe_analy})] includes  both the turbulent and viscous regimes.

Earlier Grossmann and Lohse~\cite{Grossmann:JFM2000, Grossmann:PRL2001, Grossmann:PRE2002} estimated $U$ and the Nusselt number by invoking the exact relations of Shraiman and Siggia~\cite{Shraiman:PRA1990} and using the fact that the total dissipation is a sum of those in the bulk and in the boundary layers ($D_{u,bulk}$ and $D_{u,BL}$ respectively).  Our derivation is an alternative to that of GL with an attempt to highlight the anisotropic effects arising due to the boundary layers that yield $\epsilon_u  \ne U^3/d$ and $\epsilon_T  \ne U \Delta^2 /d$. Note that our derivation  does not use Shraiman and Siggia's~\cite{Shraiman:PRA1990}  exact relations,  $\epsilon_{u,bulk}  \sim U^3/d$, and $\epsilon_{T,bulk}  \sim U \Delta^2 /d$, where $\epsilon_{u,bulk}$ and $\epsilon_{T,bulk}$ are the  bulk viscous and thermal dissipation rates respectively.

\section{Conclusions}

The agreement between our model and earlier experiments and numerical simulations is  remarkable, considering that our prediction is based on cubical box, while the experiments and numerical simulations employ cubical and cylindrical geometry.  This result indicates that the P\'{e}clet scaling is {\em weakly} dependent on the aspect ratio or geometry, and it can be reasonably well described by the nonlinear equation [Eq.~(\ref{eq:U})] which is based on the scaling of the large-scale quantities.  This is one of the important conclusions of our work as well as that of Grossmann and Lohse~\cite{Grossmann:JFM2000, Grossmann:PRL2001, Grossmann:PRE2002}. Note however that the discrepancies between the model predictions and the experiments results (see Fig.~\ref{fig:pe}) could be due to weak dependence of P\'{e}clet number on geometry or aspect ratio.

The above observations indicate that  the flow behaviour in RBC differs significantly from the unbounded hydrodynamic turbulence for which we employ homogeneous and isotropic formalism.   Interestingly, in turbulent RBC, the buoyancy term is {\em nearly} cancelled by the viscous term.  This feature of RBC could be the reason for the Kolmogorov's spectrum in RBC, as reported by Kumar {\em et al.}~\cite{Kumar:PRE2014}.  The aforementioned wall effects should also be present in other bounded flows such as in channels, pipes, rotating convection, spheres, and cylinders.  The procedure adopted in this paper would yield similar formulae for the large-scale velocity and the dissipation rate for these systems.

In summary, we derive a general formula for the large scale velocity $U$ or the P\'{e}clet number for RBC that is applicable for any $\mathrm{Ra}$ and $\mathrm{Pr}$.  Our formula provides reasonable fits to the results of earlier experiments and numerical simulations.  We also compute the correlation function between $u_z$ and $\theta$ that causes deviations of  the Nusselt number  from the theoretical prediction of $ \mathrm{Nu}  \sim \mathrm{Ra}^{1/2}$ to the experimentally observed $ \mathrm{Nu}  \sim \mathrm{Ra}^{0.30}$.  In Table~\ref{tab:Ra_dependence}, we also show how the dissipation rate $\epsilon_u$ and the temperature fluctuations in RBC get corrections from the usual formulas due to the boundary walls.

Our formulae discussed in this paper provide insights into the flow dynamics of RBC.  These results will  be useful in modelling convective flows in the interiors and atmospheres of stars and planets, as well as in engineering applications. 

\acknowledgements
The simulations were performed on the HPC system and Chaos cluster of  IIT Kanpur, India, and Shaheen-II supercomputer of KAUST, Saudi Arabia.  This work was supported by a research grant SERB/F/3279/2013-14 from Science and Engineering Research Board, India.

\section*{Appendix: Temperature profile and boundary layer}

The temperature is a function of $x,y,z$, i.e., $T(x,y,z)$.  However, as shown in Fig.~\ref{fig:Tz}, its planar average, $T_m(z)$, is approximately $1/2$ in the bulk, and it varies rapidly in the boundary layers.  We can approximate $T_m(z)$ as  
 \begin{equation}
    T_m(z)= 
\begin{cases}
    1 - \frac{z}{2\delta_T} & \text{for } 0 < z < \delta_T \\
	1/2  & \text{for } \delta_T < z < 1 - \delta_T \\
	\frac{1-z}{2\delta_T} & \text{for } 1 - \delta_T < z < 1 
\end{cases}
\end{equation}
where $\delta_T$ is the thickness of the thermal boundary layers at the top and bottom plates.  In RBC, it is customary to describe the flow using the temperature fluctuation from the conduction state, $\theta$, defined as
 \begin{equation}
T(x,y,z) = \theta(x,y,z) + 1-z .
\label{eq:theta(x,y,z)}
\end{equation}
For the above, we have normalized the temperature fluctuation using the temperature difference between the plates, and the vertical coordinate using the vertical distance between the plates.  When we perform averaging of Eq.~(\ref{eq:theta(x,y,z)}) over $xy$ planes, we obtain
\begin{equation}
\theta_m(z) = T_m(z) + z-1,
\label{eq:theta(z)}
\end{equation}
where $\theta_m(z)$ is
 \begin{equation}
    \theta_m(z) = 
\begin{cases}
    z \left( 1 - \frac{1}{2\delta_T} \right ) & \text{for } 0 < z < \delta_T \\
	z - 1/2 & \text{for } \delta_T < z < 1 - \delta_T \\
    (z-1)\left( 1 - \frac{1}{2\delta_T} \right ) & \text{for } 1 - \delta_T < z < 1 
\end{cases}
\end{equation}
which is exhibited in Fig.~\ref{fig:Tz}. For a pair of thin boundary layers, the Fourier transform of $\theta_m(z)$, $\hat{\theta}_m(k_z)$, is dominated by the contributions from the bulk, that is,
\begin{eqnarray}
\hat{\theta}_m(k_z) & = & \int_0^1 \theta_m(z) \sin(k_z \pi z) dz \nonumber \\
 & \approx & \int_0^1 (z-1/2) \sin(k_z \pi z) dz  \nonumber\\
 & \approx & \begin{cases}
       -\frac{1}{\pi k_z} &\mathrm{for \, even} \, k_z \\
        0 \, &\mathrm{otherwise}
\end{cases}
\end{eqnarray}

It is interesting to note that the corresponding velocity mode,  $\hat{u}_z(0, 0, k_z) = 0$ because of the incompressibility condition ${\bf k} \cdot \hat{\bf u}(0, 0, k_z) = k_z \hat{u}_z(0,0,k_z) = 0$.  Also, $\hat{u}_{x,y}(0, 0, k_z) = 0$ when we assume an absence of a mean horizontal flow in any horizontal plane.  Hence, the momentum equation for the Fourier mode $(0,0,k_z)$ is 
\begin{equation}
-i k_z \hat{\sigma}_m(0,0,k_z) + \alpha g \hat{\theta}_m(0,0,k_z) = 0,
\end{equation}
and it does not involve the velocity field. In the real space, the above equation translates to $d \sigma_m/dz = \alpha g \theta_m(z)$. For the Fourier modes other than $(0,0,k_z)$, the momentum equation is
\begin{eqnarray}
\frac{\partial \hat{{\bf u}}({\bf k})}{\partial t} & = &- i \sum_{\bf{p+q=k}} {[\bf k \cdot \hat{u}}({\bf q})] {\bf \hat{u}}({\bf p}) -    i {\bf k} \hat{\sigma}_\mathrm{res}({\bf k})  \nonumber \\ 
&& +\alpha g \hat{\theta}_\mathrm{res}({\bf k}) \hat{z} - \nu k^2 \hat{{\bf u}}({\bf k}). \label{eq:u_four}
\end{eqnarray}
We denote the participating temperature field in the above equation as residual temperature $\theta_\mathrm{res}$, and the residual pressure field as $\sigma_\mathrm{res}$, and they are defined as
 \begin{eqnarray}
\theta_\mathrm{res} & = & \theta - \theta_m, \\
\sigma_\mathrm{res} & = & \sigma - \sigma_m.
\label{eq:theta_res}
\end{eqnarray}
Thus, the large-scale velocity, $U$, depends on $\theta_\mathrm{res}$ and $\sigma_\mathrm{res}$.


\begin{thebibliography}{41}
\expandafter\ifx\csname natexlab\endcsname\relax\def\natexlab#1{#1}\fi
\expandafter\ifx\csname bibnamefont\endcsname\relax
  \def\bibnamefont#1{#1}\fi
\expandafter\ifx\csname bibfnamefont\endcsname\relax
  \def\bibfnamefont#1{#1}\fi
\expandafter\ifx\csname citenamefont\endcsname\relax
  \def\citenamefont#1{#1}\fi
\expandafter\ifx\csname url\endcsname\relax
  \def\url#1{\texttt{#1}}\fi
\expandafter\ifx\csname urlprefix\endcsname\relax\def\urlprefix{URL }\fi
\providecommand{\bibinfo}[2]{#2}
\providecommand{\eprint}[2][]{\url{#2}}

\bibitem[{\citenamefont{{Pope}}(2000)}]{Pope:Book}
\bibinfo{author}{\bibfnamefont{S.~B.} \bibnamefont{{Pope}}},
  \emph{\bibinfo{title}{Turbulent Flows}} (\bibinfo{publisher}{Cambridge
  University Press}, \bibinfo{address}{Cambridge}, \bibinfo{year}{2000}).

\bibitem[{\citenamefont{Davidson}(2004)}]{Davidson:book2004}
\bibinfo{author}{\bibfnamefont{P.~A.} \bibnamefont{Davidson}},
  \emph{\bibinfo{title}{Turbulence: an introduction for scientists and
  engineers}} (\bibinfo{publisher}{Oxford University Press},
  \bibinfo{address}{Oxford, UK}, \bibinfo{year}{2004}).

\bibitem[{\citenamefont{{Lesieur}}(2008)}]{Lesieur:book}
\bibinfo{author}{\bibfnamefont{M.}~\bibnamefont{{Lesieur}}},
  \emph{\bibinfo{title}{Turbulence in Fluids - Stochastic and Numerical
  Modelling}} (\bibinfo{publisher}{Kluwer Academic Publishers},
  \bibinfo{address}{Dordrecht}, \bibinfo{year}{2008}).

\bibitem[{\citenamefont{{Frisch}}(2011)}]{Frisch:Book}
\bibinfo{author}{\bibfnamefont{U.}~\bibnamefont{{Frisch}}},
  \emph{\bibinfo{title}{Turbulence: The Legacy of A N Kolmogorov}}
  (\bibinfo{publisher}{Cambridge University Press},
  \bibinfo{address}{Cambridge}, \bibinfo{year}{2011}).

\bibitem[{\citenamefont{{Davidson}}(2013)}]{Davidson:book2013}
\bibinfo{author}{\bibfnamefont{P.~A.} \bibnamefont{{Davidson}}},
  \emph{\bibinfo{title}{Turbulence in Rotating Stratified and Electrically
  Conducting Fluids}} (\bibinfo{publisher}{Cambridge University Press},
  \bibinfo{address}{Cambridge}, \bibinfo{year}{2013}).

\bibitem[{\citenamefont{{Ahlers} et~al.}(2009)\citenamefont{{Ahlers},
  {Grossmann}, and {Lohse}}}]{Ahlers:RMP2009}
\bibinfo{author}{\bibfnamefont{G.}~\bibnamefont{{Ahlers}}},
  \bibinfo{author}{\bibfnamefont{S.}~\bibnamefont{{Grossmann}}},
  \bibnamefont{and} \bibinfo{author}{\bibfnamefont{D.}~\bibnamefont{{Lohse}}},
  \bibinfo{journal}{Rev. Mod. Phys.} \textbf{\bibinfo{volume}{81}},
  \bibinfo{pages}{503} (\bibinfo{year}{2009}).

\bibitem[{\citenamefont{{Siggia}}(1994)}]{Siggia:ARFM1994}
\bibinfo{author}{\bibfnamefont{E.~D.} \bibnamefont{{Siggia}}},
  \bibinfo{journal}{Annu. Rev. Fluid Mech.} \textbf{\bibinfo{volume}{26}},
  \bibinfo{pages}{137} (\bibinfo{year}{1994}).

\bibitem[{\citenamefont{{Lohse} and {Xia}}(2010)}]{Lohse:ARFM2010}
\bibinfo{author}{\bibfnamefont{D.}~\bibnamefont{{Lohse}}} \bibnamefont{and}
  \bibinfo{author}{\bibfnamefont{K.~Q.} \bibnamefont{{Xia}}},
  \bibinfo{journal}{Annu. Rev. Fluid Mech.} \textbf{\bibinfo{volume}{42}},
  \bibinfo{pages}{335} (\bibinfo{year}{2010}).

\bibitem[{\citenamefont{{Chill\`{a}} and {Schumacher}}(2012)}]{Chilla:EPJE2012}
\bibinfo{author}{\bibfnamefont{F.}~\bibnamefont{{Chill\`{a}}}}
  \bibnamefont{and}
  \bibinfo{author}{\bibfnamefont{J.}~\bibnamefont{{Schumacher}}},
  \bibinfo{journal}{Eur. Phys. J. E} \textbf{\bibinfo{volume}{35}},
  \bibinfo{pages}{58} (\bibinfo{year}{2012}).

\bibitem[{\citenamefont{{Grossmann} and {Lohse}}(2000)}]{Grossmann:JFM2000}
\bibinfo{author}{\bibfnamefont{S.}~\bibnamefont{{Grossmann}}} \bibnamefont{and}
  \bibinfo{author}{\bibfnamefont{D.}~\bibnamefont{{Lohse}}},
  \bibinfo{journal}{J. Fluid Mech.} \textbf{\bibinfo{volume}{407}},
  \bibinfo{pages}{27} (\bibinfo{year}{2000}).

\bibitem[{\citenamefont{{Grossmann} and {Lohse}}(2001)}]{Grossmann:PRL2001}
\bibinfo{author}{\bibfnamefont{S.}~\bibnamefont{{Grossmann}}} \bibnamefont{and}
  \bibinfo{author}{\bibfnamefont{D.}~\bibnamefont{{Lohse}}},
  \bibinfo{journal}{Phys. Rev. Lett.} \textbf{\bibinfo{volume}{86}},
  \bibinfo{pages}{3316} (\bibinfo{year}{2001}).

\bibitem[{\citenamefont{{Grossmann} and {Lohse}}(2002)}]{Grossmann:PRE2002}
\bibinfo{author}{\bibfnamefont{S.}~\bibnamefont{{Grossmann}}} \bibnamefont{and}
  \bibinfo{author}{\bibfnamefont{D.}~\bibnamefont{{Lohse}}},
  \bibinfo{journal}{Phys. Rev. E} \textbf{\bibinfo{volume}{66}},
  \bibinfo{pages}{016305} (\bibinfo{year}{2002}).

\bibitem[{\citenamefont{{Stevens} et~al.}(2013)\citenamefont{{Stevens}, {Poel},
  {Grossmann}, and {Lohse}}}]{Stevens:JFM2013}
\bibinfo{author}{\bibfnamefont{R.}~\bibnamefont{{Stevens}}},
  \bibinfo{author}{\bibfnamefont{E.~P.} \bibnamefont{{Poel}}},
  \bibinfo{author}{\bibfnamefont{S.}~\bibnamefont{{Grossmann}}},
  \bibnamefont{and} \bibinfo{author}{\bibfnamefont{D.}~\bibnamefont{{Lohse}}},
  \bibinfo{journal}{J. Fluid Mech.} \textbf{\bibinfo{volume}{730}},
  \bibinfo{pages}{295} (\bibinfo{year}{2013}).

\bibitem[{\citenamefont{{Xin} and {Xia}}(1997)}]{Xin:PRE1997}
\bibinfo{author}{\bibfnamefont{Y.~B.} \bibnamefont{{Xin}}} \bibnamefont{and}
  \bibinfo{author}{\bibfnamefont{K.~Q.} \bibnamefont{{Xia}}},
  \bibinfo{journal}{Phys. Rev. E} \textbf{\bibinfo{volume}{56}},
  \bibinfo{pages}{3010} (\bibinfo{year}{1997}).

\bibitem[{\citenamefont{{Cioni} et~al.}(1997)\citenamefont{{Cioni},
  {Ciliberto}, and {Sommeria}}}]{Cioni:JFM1997}
\bibinfo{author}{\bibfnamefont{S.}~\bibnamefont{{Cioni}}},
  \bibinfo{author}{\bibfnamefont{S.}~\bibnamefont{{Ciliberto}}},
  \bibnamefont{and}
  \bibinfo{author}{\bibfnamefont{J.}~\bibnamefont{{Sommeria}}},
  \bibinfo{journal}{J. Fluid Mech.} \textbf{\bibinfo{volume}{335}},
  \bibinfo{pages}{111} (\bibinfo{year}{1997}).

\bibitem[{\citenamefont{{Qiu} and {Tong}}(2001)}]{Qiu:PRL2001}
\bibinfo{author}{\bibfnamefont{X.~L.} \bibnamefont{{Qiu}}} \bibnamefont{and}
  \bibinfo{author}{\bibfnamefont{P.}~\bibnamefont{{Tong}}},
  \bibinfo{journal}{Phys. Rev. Lett.} \textbf{\bibinfo{volume}{87}},
  \bibinfo{pages}{094501} (\bibinfo{year}{2001}).

\bibitem[{\citenamefont{{Ahlers} and {Xu}}(2001)}]{Ahlers:PRL2001}
\bibinfo{author}{\bibfnamefont{G.}~\bibnamefont{{Ahlers}}} \bibnamefont{and}
  \bibinfo{author}{\bibfnamefont{X.}~\bibnamefont{{Xu}}},
  \bibinfo{journal}{Phys. Rev. Lett.} \textbf{\bibinfo{volume}{86}},
  \bibinfo{pages}{3320} (\bibinfo{year}{2001}).

\bibitem[{\citenamefont{{Niemela} et~al.}(2001)\citenamefont{{Niemela},
  {Skrbek}, {Sreenivasan}, and {Donnelly}}}]{Niemela:JFM2001}
\bibinfo{author}{\bibfnamefont{J.~J.} \bibnamefont{{Niemela}}},
  \bibinfo{author}{\bibfnamefont{L.}~\bibnamefont{{Skrbek}}},
  \bibinfo{author}{\bibfnamefont{K.~R.} \bibnamefont{{Sreenivasan}}},
  \bibnamefont{and} \bibinfo{author}{\bibfnamefont{R.~J.}
  \bibnamefont{{Donnelly}}}, \bibinfo{journal}{J. Fluid Mech.}
  \textbf{\bibinfo{volume}{449}}, \bibinfo{pages}{169} (\bibinfo{year}{2001}).

\bibitem[{\citenamefont{{Lam} et~al.}(2002)\citenamefont{{Lam}, {Shang},
  {Zhou}, and {Xia}}}]{Lam:PRE2002}
\bibinfo{author}{\bibfnamefont{S.}~\bibnamefont{{Lam}}},
  \bibinfo{author}{\bibfnamefont{X.-D.} \bibnamefont{{Shang}}},
  \bibinfo{author}{\bibfnamefont{S.-Q.} \bibnamefont{{Zhou}}},
  \bibnamefont{and} \bibinfo{author}{\bibfnamefont{K.-Q.} \bibnamefont{{Xia}}},
  \bibinfo{journal}{Phys. Rev. E} \textbf{\bibinfo{volume}{65}},
  \bibinfo{pages}{066306} (\bibinfo{year}{2002}).

\bibitem[{\citenamefont{{Urban} et~al.}(2012)\citenamefont{{Urban}, {Hanzelka},
  {Kralik}, {Musilova}, {Srnka}, and {Skrbek}}}]{Urban:PRL2012}
\bibinfo{author}{\bibfnamefont{P.}~\bibnamefont{{Urban}}},
  \bibinfo{author}{\bibfnamefont{P.}~\bibnamefont{{Hanzelka}}},
  \bibinfo{author}{\bibfnamefont{T.}~\bibnamefont{{Kralik}}},
  \bibinfo{author}{\bibfnamefont{V.}~\bibnamefont{{Musilova}}},
  \bibinfo{author}{\bibfnamefont{A.}~\bibnamefont{{Srnka}}}, \bibnamefont{and}
  \bibinfo{author}{\bibfnamefont{L.}~\bibnamefont{{Skrbek}}},
  \bibinfo{journal}{Phys. Rev. Lett.} \textbf{\bibinfo{volume}{109}},
  \bibinfo{pages}{154301} (\bibinfo{year}{2012}).

\bibitem[{\citenamefont{{He} et~al.}(2012)\citenamefont{{He}, {Funfschilling},
  {Nobach}, {Bodenschatz}, and {Ahlers}}}]{He:PRL2012}
\bibinfo{author}{\bibfnamefont{X.}~\bibnamefont{{He}}},
  \bibinfo{author}{\bibfnamefont{D.}~\bibnamefont{{Funfschilling}}},
  \bibinfo{author}{\bibfnamefont{H.}~\bibnamefont{{Nobach}}},
  \bibinfo{author}{\bibfnamefont{E.}~\bibnamefont{{Bodenschatz}}},
  \bibnamefont{and} \bibinfo{author}{\bibfnamefont{G.}~\bibnamefont{{Ahlers}}},
  \bibinfo{journal}{Phys. Rev. Lett.} \textbf{\bibinfo{volume}{108}},
  \bibinfo{pages}{024502} (\bibinfo{year}{2012}).

\bibitem[{\citenamefont{{Camussi} and {Verzicco}}(1998)}]{Camussi:POF1998}
\bibinfo{author}{\bibfnamefont{R.}~\bibnamefont{{Camussi}}} \bibnamefont{and}
  \bibinfo{author}{\bibfnamefont{R.}~\bibnamefont{{Verzicco}}},
  \bibinfo{journal}{Phys. Fluids} \textbf{\bibinfo{volume}{10}},
  \bibinfo{pages}{516} (\bibinfo{year}{1998}).

\bibitem[{\citenamefont{{van Reeuwijk} et~al.}(2008)\citenamefont{{van
  Reeuwijk}, Jonker, and {Hanjali\'{c}}}}]{Reeuwijk:PRE2008}
\bibinfo{author}{\bibfnamefont{M.}~\bibnamefont{{van Reeuwijk}}},
  \bibinfo{author}{\bibfnamefont{H.~J.~J.} \bibnamefont{Jonker}},
  \bibnamefont{and}
  \bibinfo{author}{\bibfnamefont{K.}~\bibnamefont{{Hanjali\'{c}}}},
  \bibinfo{journal}{Phys. Rev. E} \textbf{\bibinfo{volume}{77}},
  \bibinfo{pages}{036311} (\bibinfo{year}{2008}).

\bibitem[{\citenamefont{{Silano} et~al.}(2010)\citenamefont{{Silano},
  {Sreenivasan}, and {Verzicco}}}]{Silano:JFM2010}
\bibinfo{author}{\bibfnamefont{G.}~\bibnamefont{{Silano}}},
  \bibinfo{author}{\bibfnamefont{K.~R.} \bibnamefont{{Sreenivasan}}},
  \bibnamefont{and}
  \bibinfo{author}{\bibfnamefont{R.}~\bibnamefont{{Verzicco}}},
  \bibinfo{journal}{J. Fluid Mech.} \textbf{\bibinfo{volume}{662}},
  \bibinfo{pages}{409} (\bibinfo{year}{2010}).

\bibitem[{\citenamefont{{Bailon-Cuba} et~al.}(2010)\citenamefont{{Bailon-Cuba},
  {Emran}, and {Schumacher}}}]{BailonCuba:JFM2010}
\bibinfo{author}{\bibfnamefont{J.}~\bibnamefont{{Bailon-Cuba}}},
  \bibinfo{author}{\bibfnamefont{M.~S.} \bibnamefont{{Emran}}},
  \bibnamefont{and}
  \bibinfo{author}{\bibfnamefont{J.}~\bibnamefont{{Schumacher}}},
  \bibinfo{journal}{J. Fluid Mech.} \textbf{\bibinfo{volume}{655}},
  \bibinfo{pages}{152} (\bibinfo{year}{2010}).

\bibitem[{\citenamefont{{Scheel} et~al.}(2012)\citenamefont{{Scheel}, {Kim},
  and {White}}}]{Scheel:JFM2012}
\bibinfo{author}{\bibfnamefont{J.~D.} \bibnamefont{{Scheel}}},
  \bibinfo{author}{\bibfnamefont{E.}~\bibnamefont{{Kim}}}, \bibnamefont{and}
  \bibinfo{author}{\bibfnamefont{K.~R.} \bibnamefont{{White}}},
  \bibinfo{journal}{J. Fluid Mech.} \textbf{\bibinfo{volume}{711}},
  \bibinfo{pages}{281} (\bibinfo{year}{2012}).

\bibitem[{\citenamefont{{Verma} et~al.}(2012)\citenamefont{{Verma}, {Mishra},
  {Pandey}, and {Paul}}}]{Verma:PRE2012}
\bibinfo{author}{\bibfnamefont{M.~K.} \bibnamefont{{Verma}}},
  \bibinfo{author}{\bibfnamefont{P.~K.} \bibnamefont{{Mishra}}},
  \bibinfo{author}{\bibfnamefont{A.}~\bibnamefont{{Pandey}}}, \bibnamefont{and}
  \bibinfo{author}{\bibfnamefont{S.}~\bibnamefont{{Paul}}},
  \bibinfo{journal}{Phys. Rev. E} \textbf{\bibinfo{volume}{85}},
  \bibinfo{pages}{016310} (\bibinfo{year}{2012}).

\bibitem[{\citenamefont{{Wagner} and {Shishkina}}(2013)}]{Wagner:POF2013}
\bibinfo{author}{\bibfnamefont{S.}~\bibnamefont{{Wagner}}} \bibnamefont{and}
  \bibinfo{author}{\bibfnamefont{O.}~\bibnamefont{{Shishkina}}},
  \bibinfo{journal}{Phys. Fluids} \textbf{\bibinfo{volume}{25}},
  \bibinfo{pages}{085110} (\bibinfo{year}{2013}).

\bibitem[{\citenamefont{{Pandey} et~al.}(2014)\citenamefont{{Pandey}, {Verma},
  and {Mishra}}}]{Pandey:PRE2014}
\bibinfo{author}{\bibfnamefont{A.}~\bibnamefont{{Pandey}}},
  \bibinfo{author}{\bibfnamefont{M.~K.} \bibnamefont{{Verma}}},
  \bibnamefont{and} \bibinfo{author}{\bibfnamefont{P.~K.}
  \bibnamefont{{Mishra}}}, \bibinfo{journal}{Phys. Rev. E}
  \textbf{\bibinfo{volume}{89}}, \bibinfo{pages}{023006}
  (\bibinfo{year}{2014}).

\bibitem[{\citenamefont{{Horn} et~al.}(2013)\citenamefont{{Horn}, {Shishkina},
  and {Wagner}}}]{Horn:JFM2013}
\bibinfo{author}{\bibfnamefont{S.}~\bibnamefont{{Horn}}},
  \bibinfo{author}{\bibfnamefont{O.}~\bibnamefont{{Shishkina}}},
  \bibnamefont{and} \bibinfo{author}{\bibfnamefont{C.}~\bibnamefont{{Wagner}}},
  \bibinfo{journal}{J. Fluid Mech.} \textbf{\bibinfo{volume}{724}},
  \bibinfo{pages}{175} (\bibinfo{year}{2013}).

\bibitem[{\citenamefont{OpenFOAM}(2015)}]{OpenFOAM}
\bibinfo{author}{\bibnamefont{OpenFOAM}}, \emph{\bibinfo{title}{Openfoam: The
  open source cfd toolbox}} (\bibinfo{year}{2015}),
  \urlprefix\url{http://www.openfoam.org}.

\bibitem[{\citenamefont{{Gr\"otzbach}}(1983)}]{Grotzbach:JCP1983}
\bibinfo{author}{\bibfnamefont{G.}~\bibnamefont{{Gr\"otzbach}}},
  \bibinfo{journal}{J. Comp. Phys.} \textbf{\bibinfo{volume}{49}},
  \bibinfo{pages}{241} (\bibinfo{year}{1983}).

\bibitem[{\citenamefont{{Shishkina} et~al.}(2010)\citenamefont{{Shishkina},
  {Stevens}, {Grossmann}, and {Lohse}}}]{Shishkina:NJP2010}
\bibinfo{author}{\bibfnamefont{O.}~\bibnamefont{{Shishkina}}},
  \bibinfo{author}{\bibfnamefont{R.}~\bibnamefont{{Stevens}}},
  \bibinfo{author}{\bibfnamefont{S.}~\bibnamefont{{Grossmann}}},
  \bibnamefont{and} \bibinfo{author}{\bibfnamefont{D.}~\bibnamefont{{Lohse}}},
  \bibinfo{journal}{New J. Phys.} \textbf{\bibinfo{volume}{12}},
  \bibinfo{pages}{075022} (\bibinfo{year}{2010}).

\bibitem[{\citenamefont{{Scheel} and {Schumacher}}(2014)}]{Scheel:JFM2014}
\bibinfo{author}{\bibfnamefont{J.~D.} \bibnamefont{{Scheel}}} \bibnamefont{and}
  \bibinfo{author}{\bibfnamefont{J.}~\bibnamefont{{Schumacher}}},
  \bibinfo{journal}{J. Fluid Mech.} \textbf{\bibinfo{volume}{758}},
  \bibinfo{pages}{373} (\bibinfo{year}{2014}).

\bibitem[{\citenamefont{{Schubert} et~al.}(2001)\citenamefont{{Schubert},
  {Turcotte}, and {Olson}}}]{Schubert:book2001}
\bibinfo{author}{\bibfnamefont{G.}~\bibnamefont{{Schubert}}},
  \bibinfo{author}{\bibfnamefont{D.~L.} \bibnamefont{{Turcotte}}},
  \bibnamefont{and} \bibinfo{author}{\bibfnamefont{P.}~\bibnamefont{{Olson}}},
  \emph{\bibinfo{title}{Mantle Convection in the Earth and Planets}}
  (\bibinfo{publisher}{Cambridge University Press},
  \bibinfo{address}{Cambridge, UK}, \bibinfo{year}{2001}).

\bibitem[{\citenamefont{{Turcotte} and {Schubert}}(2002)}]{Turcotte:book2002}
\bibinfo{author}{\bibfnamefont{D.~L.} \bibnamefont{{Turcotte}}}
  \bibnamefont{and}
  \bibinfo{author}{\bibfnamefont{G.}~\bibnamefont{{Schubert}}},
  \emph{\bibinfo{title}{Geodynamics}} (\bibinfo{publisher}{Cambridge University
  Press}, \bibinfo{address}{Cambridge, UK}, \bibinfo{year}{2002}).

\bibitem[{\citenamefont{{Galsa} et~al.}(2015)\citenamefont{{Galsa}, {Herein},
  {Lenkey}, {Farkas}, and {Taller}}}]{Galsa:SE2015}
\bibinfo{author}{\bibfnamefont{A.}~\bibnamefont{{Galsa}}},
  \bibinfo{author}{\bibfnamefont{M.}~\bibnamefont{{Herein}}},
  \bibinfo{author}{\bibfnamefont{L.}~\bibnamefont{{Lenkey}}},
  \bibinfo{author}{\bibfnamefont{M.~P.} \bibnamefont{{Farkas}}},
  \bibnamefont{and} \bibinfo{author}{\bibfnamefont{G.}~\bibnamefont{{Taller}}},
  \bibinfo{journal}{Solid Earth} \textbf{\bibinfo{volume}{6}},
  \bibinfo{pages}{93} (\bibinfo{year}{2015}).

\bibitem[{\citenamefont{{Kraichnan}}(1962)}]{Kraichnan:POF1962}
\bibinfo{author}{\bibfnamefont{R.~H.} \bibnamefont{{Kraichnan}}},
  \bibinfo{journal}{Phys. Fluids} \textbf{\bibinfo{volume}{5}},
  \bibinfo{pages}{1374} (\bibinfo{year}{1962}).

\bibitem[{\citenamefont{{Shraiman} and {Siggia}}(1990)}]{Shraiman:PRA1990}
\bibinfo{author}{\bibfnamefont{B.~I.} \bibnamefont{{Shraiman}}}
  \bibnamefont{and} \bibinfo{author}{\bibfnamefont{E.~D.}
  \bibnamefont{{Siggia}}}, \bibinfo{journal}{Phys. Rev. A}
  \textbf{\bibinfo{volume}{42}}, \bibinfo{pages}{3650} (\bibinfo{year}{1990}).

\bibitem[{\citenamefont{{Landau} and {Lifshitz}}(1987)}]{Landau:FM1987}
\bibinfo{author}{\bibfnamefont{L.~D.} \bibnamefont{{Landau}}} \bibnamefont{and}
  \bibinfo{author}{\bibfnamefont{E.~M.} \bibnamefont{{Lifshitz}}},
  \emph{\bibinfo{title}{Fluid Mechanics}} (\bibinfo{publisher}{Pergamon},
  \bibinfo{address}{Oxford}, \bibinfo{year}{1987}).

\bibitem[{\citenamefont{{Kumar} et~al.}(2014)\citenamefont{{Kumar},
  {Chatterjee}, and {Verma}}}]{Kumar:PRE2014}
\bibinfo{author}{\bibfnamefont{A.}~\bibnamefont{{Kumar}}},
  \bibinfo{author}{\bibfnamefont{A.~G.} \bibnamefont{{Chatterjee}}},
  \bibnamefont{and} \bibinfo{author}{\bibfnamefont{M.~K.}
  \bibnamefont{{Verma}}}, \bibinfo{journal}{Phys. Rev. E}
  \textbf{\bibinfo{volume}{90}}, \bibinfo{pages}{023016}
  (\bibinfo{year}{2014}).

\end{thebibliography}
\end{document}